\documentclass[letterpaper, 10 pt, conference]{ieeeconf}  

\IEEEoverridecommandlockouts                              

\usepackage{cite}
\usepackage{balance}
\usepackage{amsthm}
\usepackage{amssymb}
\newtheorem{Theorem}{Theorem}[]
\newtheorem{Definition}{Definition}[]
\newtheorem{Assumption}{Assumption}[]
\usepackage{xcolor,soul,framed} 
\usepackage[pdftex]{graphicx}
\usepackage[cmex10]{amsmath}
\usepackage{array}
\usepackage{algorithm}
\usepackage{algpseudocode}
\usepackage{amsfonts}
\usepackage{tabularx} 
\usepackage{amsfonts} 
\usepackage{hyperref}

\usepackage{algorithmicx}
\algnewcommand\algorithmicforeach{\textbf{for each}}
\algdef{S}[FOR]{ForEach}[1]{\algorithmicforeach\ #1\ \algorithmicdo}

\title{\LARGE \bf
Data-Driven Safety Filter: An Input-Output Perspective
}

\author{Mohammad Bajelani and Klaske van Heusden
\thanks{We acknowledge the support of the Natural Sciences and Engineering Research Council of Canada (NSERC) [RGPIN-2023-03660].}
\thanks{Mohammad Bajelani and Klaske van Heusden are with the University of British Columbia, School of Engineering, 3333 University Way, Kelowna, BC V1V 1V7
{\tt\small mohammad.bajelani, klaske.vanheusden @ubc.ca}}}

\begin{document}
\maketitle
\thispagestyle{empty}
\pagestyle{empty}

\begin{abstract}

Implementation of learning-based control remains challenging due to the absence of safety guarantees. Safe control methods have turned to model-based safety filters to address these challenges, but this is paradoxical when the ultimate goal is a model-free, data-driven control solution. Addressing the core question of ``Can we ensure the safety of any learning-based algorithm without explicit prediction models and state estimation?'' this paper proposes a Data-Driven Safety Filter (DDSF) grounded in Behavioral System Theory (BST). The proposed method needs only a single system trajectory available in an offline dataset to modify unsafe learning inputs to safe inputs. This contribution addresses safe control in the input-output framework and therefore does not require full state measurements or explicit state estimation. Since no explicit model is required, the proposed safe control solution is not affected by unmodeled dynamics and unstructured uncertainty and can provide a safe solution for systems with unknown time delays. The effectiveness of the proposed DDSF is illustrated in simulation for a high-order six-degree-of-freedom aerial robot and a time-delay adaptive cruise control system. 

\end{abstract}


\section{Introduction}

Autonomous systems have become increasingly common in recent years, spurring research to ensure these control systems are safe. Methods based on control theory that can provide a safe solution typically rely on modeling approaches that demand extensive expertise, time, and effort. Conversely, learning-based methods, like Reinforcement Learning (RL), represent a more general approach that depends on input-output data obtained through extensive trial and error. Despite their ability to handle complex and uncertain tasks without a system model, learning-based controls often lack sufficient safety guarantees, making them risky for real-world applications. To mitigate these risks, the control community has introduced safe control methods that can be implemented as modular add-on safety filters, designed to enhance any learning-based controller. For a comprehensive overview of safe learning-based controllers, see \cite{brunke2022safe}. However, state-of-the-art safety filters are heavily reliant on system models and developed in the state-space framework. This paper proposes a purely data-driven input-output safe control solution. 

Safety is commonly defined as a set of permissible input-output constraints. Various methods such as Safety Preserving Control \cite{7318397}, Explicit Reference Governors \cite{8796241}, Control Barrier Functions \cite{kiss2023control}, and Model Predictive Control (MPC) \cite{wabersich2018linear} can be employed to ensure the system remains within this permissible set.

Safety filters have been introduced as a modular add-on for a learning controller. These filters ensure that the system is safe by minimally altering potentially unsafe learning inputs, regardless of the learning algorithm in use. Safety filters serve as a bridge between control theory and a diverse range of learning algorithms, from Reinforcement Learning (RL) to Human-in-the-Loop (HITL). As a result, they enable the realization of a more generalized safe learning process.

There are three main perspectives to designing such a filter, also known as safety certification. Inspired by model predictive control, Model Predictive Safety Filters (MPSF) determine the safety of the learning input at each time step by constructing a backup trajectory to a terminal safe set \cite{wabersich2018linear, wabersich2021predictive}. Hamilton-Jacobi reachability analysis proposes a general method to compute the reachable safe set even in the presence of disturbances by solving partial differential equations. Although it provides a general solution, its scalability is still a work in progress, with advancements being made through system decomposition \cite{herbert2021scalable}. Control Barrier Functions (CBFs) assure system safety by ensuring a predefined safe set is invariant. However, the synthesis of barrier functions is case-dependent, and a general method to accomplish this is currently being developed \cite{zhao2020synthesizing}. For a broader overview of these topics, refer to \cite{8796030,chen2018hamilton,hewing2020learning}.

\begin{figure} [t]
  \centering
      \includegraphics[width=0.49\textwidth]{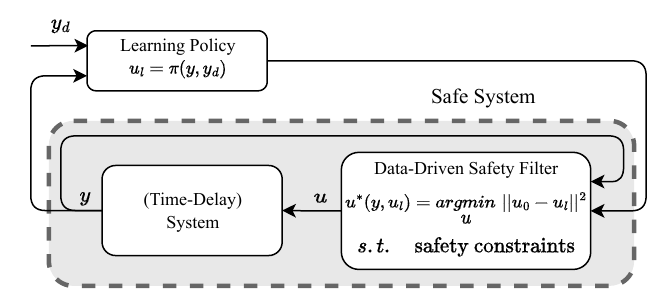}
      \caption{General block diagram of the proposed Data-Driven Safety Filter (DDSF) illustrating its application in the (time-delay) input-output framework.}
      \label{fig: DDSF_general}
      \vspace{-0.5cm}
\end{figure}

The primary limitation of these methods is their dependency on the system's model, which can be addressed through strategies such as first-principle modeling, system identification, or machine learning. Developed in the state-space framework, these methods tend to be conservative due to their sequential strategy in defining uncertainty in the predictions, i.e., defining the prediction error is more straightforward with multi-step predictors \cite{kohler2022state}. It should be noted that any model assumption made in the identification process is associated with a model mismatch, even in linear time-invariant systems \cite{martin2023guarantees}. To tackle these challenges, we propose a purely data-driven approach that ensures the system is inherently safe based solely on input-output measurements. The general block diagram of this data-driven safety filter defined in the input-output framework is illustrated in Fig.~\ref{fig: DDSF_general}.

By leveraging this method, we convert unsafe trajectories into safe ones, independent of any explicit models, thereby providing a purely data-driven learning framework. Inspired by MPSF \cite{wabersich2018linear}, we propose a data-driven safety-filter perspective in the {input-output} framework. Below, we outline the key contributions of this paper:
\begin{itemize}
    \item An input-output safe control solution: Our proposed data-driven safety filter requires only input-output measurements. No full state measurement or explicit state estimation is needed. 
    \item An entirely data-driven safe control solution: Our proposed solution eliminates the need for explicit models. 
    \item A safe control solution that is not affected by unmodeled dynamics and unstructured uncertainty. 
    \item A safe control solution that is applicable to unknown time delays. 
\end{itemize}

Note that no explicit model is defined in the data-driven control concept. In the behavioral system theory, predictions are generated based on an implicit model for which, in some instances, an equivalent data-driven explicit model can be defined \cite{10148057,9654975}. Whether an implicit or explicit model is used, the proposed framework extends MPSF \cite{wabersich2018linear} to the input-output framework and allows for safe control of systems with unknown time delays, which is incredibly challenging in the state-space framework. Note that for readability, the formulation in this paper is derived for deterministic discrete-time LTI systems; however, it is possible to extend this methodology by using robust \cite{strasser2023robust} and nonlinear \cite{berberich2020trajectory} versions of data-driven predictive controllers.
 
The remainder of the paper is organized as follows. Section II describes model-based safety filters in the state-space framework and behavioral system theory. Safety in the input-output framework is defined in Section III, along with the proposed DDSF and proof of safety. In section IV, simulation examples illustrate the performance of DDSF. Lastly, section V contains a discussion and concluding remarks.

\section{Preliminary Material} \label{Preliminary Materials}

Consider a dynamical system represented by its minimal discrete-time state-space form as follows:
\begin{equation}  \label{Discrete State Space}
    \begin{matrix}
        x(t+1) = A x(t)+B u(t), \\
        y(t) = C x(t) +D u(t),
    \end{matrix}    
\end{equation}
where $A \in \mathbb{R}^{n \times n}, B \in \mathbb{R}^{n \times m}, C \in \mathbb{R}^{p \times n}, D \in \mathbb{R}^{p \times m} $, and $x(t) \in \mathbb{R}^n, u(t) \in \mathbb{R}^m, y(t) \in \mathbb{R}^p$ are respectively the states, control inputs, and outputs at time $t \in \mathbb{Z}_{\geq 0}$, where $\mathbb{R}$ and $\mathbb{Z}_{\geq 0}$ are respectively the real numbers and non-negative integers. Additionally, this system is subjected to an initial condition $x(0)=x_0$, as well as constraints related to inputs $\mathcal{U} \subseteq \mathbb{R}^m$, states $\mathcal{X} \subseteq \mathbb{R}^n$, and outputs $\mathcal{Y} \subseteq \mathbb{R}^p$, which can be expressed as linear inequalities as follows,
\begin{equation}  \label{linear inequalities}
    \begin{split}
        \mathcal{U} &:= \{ u \in \mathbb{R}^m \mid A_u u < b_u, A_u \in \mathbb{R}^{n_u \times m}, b_u \in \mathbb{R}^{n_u} \}, \\
        \mathcal{X} &:= \{ x \in \mathbb{R}^n \mid A_x x < b_x, A_x \in \mathbb{R}^{n_x \times n}, b_x \in \mathbb{R}^{n_x}\}, \\
        \mathcal{Y} &:= \{ y \in \mathbb{R}^p \mid A_y y < b_y, A_y \in \mathbb{R}^{n_p \times p}, b_y \in \mathbb{R}^{n_p}\},
    \end{split}    
\end{equation}
where $n_u,n_x,n_p$ are the number of constraints on inputs, states, and outputs.
\begin{Definition}[Safe set, state-space formulation] \label{Def: safety}
    The safe set for system (\ref{Discrete State Space}) is the set of initial conditions for which there exists an input sequence (backup trajectory) in $\mathcal{U}$ that can keep the system's state and output for an infinite time in $\mathcal{X}$ and $\mathcal{Y}$, respectively.   
\end{Definition}

The primary objective of a safety filter is to modify any unsafe learning inputs, $u_{l} \in \mathbb{R}^m $, as minimally as possible while ensuring that the system (\ref{Discrete State Space}) remains within the defined bounds. Note that in this commonly used safe set formulation, safety is defined with respect to a set of \emph{states}. In the input-output framework, these underlying states are not accessible and the order of the system is not necessarily known (i.e. the number of states is uncertain). In this paper, we extend MPSF to the input-output framework, i.e., we aim to keep the \emph{output trajectories} safe in the sense of definition (\ref{Def: safety}) without having access to states. In this input-output framework, only the constraints $\mathcal{U}$ and $\mathcal{Y}$ are considered. In the following, background on model-based safety filters in the state-space framework and behavioral system theory are provided. 

\subsection{{Model-based safety filter in state-space framework}} \label{Safety Filter: General Formulation}

Given $\mathcal{U}$ and $\mathcal{X}$, a safety filter modifies any unsafe learning control input $u_{l} \in \mathbb{R}^m $ as little as possible to keep the system within the constraints $\mathcal{X}$ for all $t \in \mathbb{Z}_{\geq 0}$. A safety control law can be formulated as a constrained optimization problem as follows,
\begin{subequations} \label{Safety filter: general}
\begin{align}
        &\underset{u_{[0,{N-1}]}}{\operatorname{min}}  \, || u_0(t)
        -u_l(t) ||_{R}^{2} \\
        \quad s.t.  & \quad {x}_{k+1} = {A} {x}_{k}+{B} u_{k}, \\
       & \quad {x}_0(t) = {x}_0, \\
        & \quad x_k(t) \in \mathcal{X}  \quad \forall k \in \{0, \ldots, N\},\\
       & \quad u_k(t) \in \mathcal{U}  \quad \forall k \in \{0, \ldots, N-1\},
\end{align}
\end{subequations}
with $N \xrightarrow{}{} \infty $. Furthermore, $||u||^2_R = u^\top R u$ denotes a quadratic cost function weighted by a positive definite matrix $R = R^\top \succ 0\in \mathbb{R}^{m \times m}$, ${x}_0$ is the system's state at time $t$, and $u_{[0,{N-1}]} = [{u_0}^\top, \hdots, {u_{N-1}}^\top]^\top$ is the vector of decision variables. In other words, the safety filter must not only ensure the learning process is safe, but also that inputs are changed as little as possible to minimize any impact of the filter on the learning process. Solving this problem requires the knowledge of the system in terms of the system's order $n$, estimated parameters $(\hat{A},\hat{B})$, and requires measuring or observing states $\hat{x}_0$. This problem cannot be solved in real time for $N \xrightarrow{}{} \infty $, and it is necessary to approximate this infinite horizon problem by a sufficiently large horizon \cite{boccia2014stability}, or a short horizon with a terminal constraint $X_T$ \cite{CHEN19981205}. Given a finite prediction horizon of length $N_p$, and a control invariant set $X_T$ (i.e., $X_T$ is an equilibrium point of the system or a set of states for which the system can be kept safe for an infinite time after $N_p$ steps) recursive feasibility of this solvable finite-time optimization problem can be shown. The safe policy given by a Receding Horizon Control (RHC) law is determined by considering the first element of the solution of the problem (\ref{Safety filter: general}) denoted by $u^{\ast}_{[0,{N-1}]} = [{u^{\ast}_0}^\top, \hdots, {u^{\ast}_{N-1}}^\top]^\top$ in each time step as follows,
\begin{equation} \label{Safety filter: RHC}
    \mu(x,u_l) = u^{\ast}_0,
\end{equation}
where $\mu$ is a safe policy that makes the system (\ref{Discrete State Space}) inherently safe regardless of the learning input $u_l$.

\begin{Definition} [Backup Trajectory]
    A safety filter's backup trajectory is defined as a trajectory provided by the solution of finite horizon approximation of problem (\ref{Safety filter: general}) from ${x}_{0}(t) = x_{0}$ to $X_T$ with the duration of $N_p$ steps (where $N_p$ is the prediction horizon) satisfying system's constraints (\ref{linear inequalities}).
\end{Definition}

We will extend these concepts of safety filters to the data-driven input-output framework in section \ref{Section: Data-Driven Safety Filter}, with the help of behavioral system theory described in \ref{Behavioral System Theory}.

\subsection{Behavioral System Theory} \label{Behavioral System Theory}

Consider a sequence of inputs (or outputs) with length of $N_0$, as $u_{[0,{N_0-1}]} = [{u_0}^\top, \hdots, {u_{N_0-1}}^\top]^\top$ (or $y_{[0,{N_0-1}]} = [{y_0}^\top, \hdots, {y_{N_0-1}}^\top]^\top$). Note that for Multi-Input Multi-Output (MIMO) systems, $u_{[0,{N_0-1}]}$ (or $y_{[0,{N_0-1}]}$) represents a vector involving all inputs (or outputs). The Hankel matrices for input data $H_L(u) \in \mathbb{R}^{(mL)\times(N_0-L+1)}$ and output data $H_L(y) \in \mathbb{R}^{(pL)\times(N_0-L+1)}$) are respectively described as follows,
\begin{subequations}
    \begin{align}
        H_L(u) &= \begin{bmatrix}
            u_0 & u_1 & \ldots & u_{N_0-L} \\
            u_1 & u_2 & \cdots & u_{N_0-L+1} \\
            \vdots & \vdots & \ddots & \vdots \\
            u_{L-1} & u_L & \ldots & u_{N_0-1}
        \end{bmatrix},\\
        H_L(y) &= \begin{bmatrix}
            y_0 & y_1 & \ldots & y_{N_0-L} \\
            y_1 & y_2 & \cdots & y_{N_0-L+1} \\
            \vdots & \vdots & \ddots & \vdots \\
            y_{L-1} & y_L & \ldots & y_{N_0-1}
        \end{bmatrix}.
    \end{align}
\end{subequations}
In contrast to classical system theory, which views systems as models with specified structures and parameters, behavioral system theory views systems through their trajectories defined in the signal subspace \cite{markovsky2006exact}. In other words, it is assumed that the $A$, $B$, $C$, and $D$ matrices' values and dimensions in equation (\ref{Discrete State Space}) are unknown. Instead, it is assumed that it is possible to access a sufficiently high order persistently exciting, noise-free input-output single trajectory of this system as described below.

\begin{Definition}[Persistently Excitation \cite{berberich2020robust}] \label{Def: PE condition}
     Let the Hankel matrix's rank be $rank(H_L(u)) = mL$, then $u \in \mathbb{R}^m$ represents a persistently exciting signal of order $L$.
\end{Definition}

\begin{Definition}[LTI System's Trajectory \cite{9654975}] \label{Def: LTI system's trajectory}
     Let $G$ be an LTI system and $(A,B,C,D)$ its minimal realization, then $\{u_k,y_k\}_{k=0}^{k=N-1}$ is an input-output sequence of this system if there exists an initial condition ${x}_{0} \in \mathbb{R}^n$ and a state sequence $\{x_k\}_{k=0}^{k=N}$ such that 
\begin{equation}
    \begin{aligned}
    &x(k+1) =A x (k)+B u (k),\\
    &y(k)=C x(k)+D u(k) \, \, \forall k \in \{0,1,2,\hdots,N-1\}.
    \end{aligned}
\end{equation}

\end{Definition}

\begin{Definition}[System's Lag \cite{9654975}] \label{definition: lag}
    $l(A, C)$ denotes the lag of the system (\ref{Discrete State Space}), which is the smallest integer that can make the observability matrix full rank.
        \begin{equation}
        O_l(A, C):=\left(C, C A, \ldots, C A^{l-1}\right).
        \end{equation}    
\end{Definition}

The following result, known as the fundamental lemma introduced by Jan Willems \cite{willems2005note}, shows that if we have access to a single finite trajectory of an LTI system and the input is persistently exciting, then all the trajectories can be parameterized by the linear combination of the columns of Hankel matrix. This theory allows us to predict unsafe behaviors and design safety filters directly based on data without needing a parametric model. For a comprehensive overview, see \cite{markovsky2021behavioral}.

\begin{Theorem}[Fundamental Lemma \cite{berberich2020robust}] \label{Fundamental Lemma}
    Let $u^d$ be persistently exciting of order $L+n$, and ${\{u^d_k},{y^d_k\}}_{k=0}^{k=N_{0}-1}$ a trajectory of $G$. Then, ${\{\bar{u}_k},{{\bar{y}_k}\}}_{k=0}^{k=N_{0}-1}$ is a trajectory of $G$ if and only if there exists $\alpha \in \mathbb{R}^{N_{0}-L+1}$ such that
\begin{equation} \label{BST model}
    \left[\begin{array}{l}
    H_L\left(u^d\right) \\
    H_L\left(y^d\right)
    \end{array}\right] \alpha=\left[\begin{array}{l}
    \bar{u} \\
    \bar{y}
    \end{array}\right] .  
\end{equation}
\end{Theorem}

It should be mentioned that irrespective of the various representations of the pre-recorded dataset, whether it is through a Hankel matrix \cite{willems2005note}, a page matrix \cite{coulson2021distributionally}, or a collection of experiments \cite{9654975}, the space of trajectories can be spanned by the pre-recorded sequences as long as the persistent excitation assumption is met. This means that one could employ either a single trajectory or multiple trajectories, extracting many random parts of length $L$ to effectively span the trajectory space. 

We divide the Hankel matrices into two parts, where the first $T_{\text{ini}}$ rows represent past data (used to fix the initial condition), and the rest represent future data (used to create the backup trajectory). The length of the past data $T_{\text{ini}}$ implicitly determines the initial condition. As there is no definition for underlying states in the behavioral framework, we need to consider a sufficiently long segment of past data to determine the behavior of the output, where $T_{\text{ini}} \geq l(A, C)$, see \cite[lemma 1]{markovsky2008data}. Similar to \cite{berberich2020robust}, we define the equilibrium point for the input-output framework with $ T_{\text{ini}} \geq l(A, C)$,

\begin{Definition} [Equilibrium Point.] \label{def: Equilibrium point.}
    If the sequence $\left\{{u}_k, {y}_k\right\}_{k=0}^{k=T_{\text{ini}}-1}$ with $\left({u}_k, {y}_k\right)=\left(u^s, y^s\right)$ for all $k \in \{0, \hdots, T_{\text{ini}}-1\}$ is a trajectory of $G$, and $ T_{\text{ini}} \geq l(A, C)$, then $\left(u^s, y^s\right) \in$ $\mathbb{R}^{m+p}$ is an equilibrium point of system (\ref{Discrete State Space}). 
\end{Definition} 

We denote the pre-recorded single-trajectory input-output dataset by ${\{u_k^d},{y_k^d\}}_{k=0}^{k=N_0-1}$ generated by system (\ref{Discrete State Space}). Furthermore, we denote decision variables by ${\{\bar{u}_k},{{\bar{y}_k}\}}_{k=-T_{\text{ini}}}^{k=N-1}$. Note that the first $T_{\text{ini}} $ elements of this sequence are used to fix the initial condition using past input-output measured data denoted by ${u}_{[t-T_{\text{ini}},-1]}$ and ${y}_{[t-T_{\text{ini}},-1]}$, and the remaining elements are used to create the safe backup trajectory.

\section{Data-Driven Safety Filter} \label{Section: Data-Driven Safety Filter}

This section introduces an input-output safety filter using data-driven prediction. We redefine the safe set and safety filter in the context of input-output measurements. Additionally, we present the Data-Driven Safety Filter (DDSF) as an optimization problem and prove its recursive feasibility. This proof effectively shows that DDSF ensures safety in the input-output framework. While the primary focus of this paper is on data-driven prediction, the definitions and formulations are grounded in input-output measurements. Consequently, these definitions can be equivalently applied to designing an input-output safety filter using an explicit model and corresponding predictions. 

\subsection{Input-Output Safe Set and Safe Initial Trajectory}

Similar to the definition of safety (\ref{Def: safety}) in the state-space framework, we define the input-output safe set as a set of \emph{safe initial trajectories}. Starting from these initial trajectories, the system's output can be maintained within a pre-specified range for all times using a sequence of admissible inputs. Note that in the input-output framework, only input-output constraints $\mathcal{U} \times \mathcal{Y}$ are considered, and there is no concern for state constraints $\mathcal{X}$. 

\begin{Definition}[Input-Output Safe Set and Safe Initial Trajectories] \label{Safe: set} 
Let ${Traj}_{_{\text{ini}}}:= \{u_k,y_k\}_{k=-T_{\text{ini}}}^{k=-1}$ and ${Traj}_{b}:= \{u_k,y_k\}_{k=0}^{k\rightarrow \infty}$ be initial and infinite-length trajectories for system (\ref{Discrete State Space}) at $k=0$. Then, ${Traj}_{_{\text{ini}}}$ and ${Traj}_{b}$ are a safe initial trajectory and a backup trajectory if both of them are in $\mathcal{U} \times \mathcal{Y}$ and can be patched, i.e., the combined trajectory ($T_{\text{ini}}$ and ${Traj}_{b}$) is also a trajectory of (1). The input-output safe set, $S \in \mathcal{U} \times \mathcal{Y}$, is defined as a set of all safe initial trajectories.
\end{Definition}

\begin{figure}[ht] 
  \centering
  \includegraphics[width=0.48\textwidth]{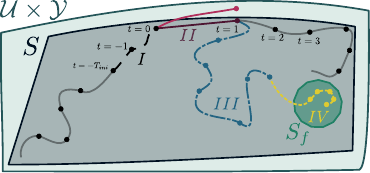}
  \caption{ Let trajectory ($I$) be an initial trajectory. At $t=0$, the proposed learning input would result in a constraint violation. DDSF will predict and modify this unsafe input as shown by segment ($II$) and generate a backup trajectory in the defined safe set, ($III$), as well as a terminal safe segment ($IV$) to keep the system's output safe.}
  \label{fig: safe set output}
\end{figure}

Calculating such an infinite-length backup trajectory is not possible. Therefore we use a terminal safe set, $S_{f}$, to truncate the tail of ${Traj}_{b}$. In other words, we will find a finite-length backup trajectory consisting of two parts, a prediction ${Traj}_{p}:= \{\bar{u}_k,\bar{y}_k\}_{k=0}^{k=N_p-1} \in \mathcal{U} \times \mathcal{Y}$, and a terminal safe trajectory, ${Traj}_s:= \{\bar{u}_k,\bar{y}_k\}_{k=N_p}^{k=N_p+T_{\text{ini}}-1} \in {S}_f$. The prediction part helps us to find a trajectory from the safe initial trajectory to the terminal safe set, and the terminal safe trajectory fixes the underlying state in the terminal safe set. This is possible due to the fact that $T_{\text{ini}}$ steps are sufficient to fix the system's state, whether in the initial condition or tail of the prediction trajectory \cite{markovsky2008data}. As a result, we will adopt the following assumptions regarding the lower bound for the prediction horizon, order of excitation, and invariant property of the terminal safe set. Fig. \ref{fig: safe set output}. provides an example of the defined terminal safe set, an unsafe learning input, and a finite-time backup trajectory.

\begin{Assumption}[Prediction Horizon Length] \label{Assumption: I}
    The prediction horizon $N_p$ is greater than $ T_{\text{ini}} \geq l(A, C)$.
\end{Assumption}
\begin{Assumption}[Persistent Excitation] \label{Assumption: II}
    The stacked Hankel matrix $[H_L(u^d) H_L(y^d)]$ is PE of order $L = N_p+2T_{\text{ini}}$ in the sense of definition (\ref{Def: PE condition}).
\end{Assumption}
\begin{Assumption}[Terminal Safe Set] \label{Assumption: III}
    The terminal set, $S_f$, is a control invariant set. The equilibrium point defined in the definition (\ref{def: Equilibrium point.}) is an obvious choice of such an invariant set.
\end{Assumption}

Given the safe set and terminal safe set as previously defined, we proceed to formulate the Data-Driven Safety Filter (DDSF) problem that employs a Hankel matrix to modify unsafe trajectories. To guarantee safety, we define an optimization problem that generates a backup trajectory ${Traj}_{p}$ that patches ${Traj}_{{\text{ini}}}$  and ${Traj}_{s}$ while modifying an unsafe input as little as possible at each time step.

\subsection{Data-Driven Safety Filter - Optimization Problem}

For an arbitrary learning input $u_l(t)$ at time $t$ and initial trajectory ${Traj}_{_{\text{ini}} }= \{u_k,y_k\}_{k=-T_{\text{ini}}}^{k=-1}$, a ${Traj}_{P}= \{\bar{u}_k,\bar{y}_k\}_{k=0}^{k={N_p-1}}$ that patches ${Traj}_{_{\text{ini}}}$ to ${Traj}_s= \{\bar{u}_k,\bar{y}_k\}_{k=N_p}^{k={N_p+T_{\text{ini}}}}$ is generated by the following quadratic optimization problem.
\begin{subequations} \label{equation: safety filter optimization problem}
\begin{align}
    &\underset{{\alpha(t)}, \bar{u}(t), \bar{y}(t)}{\operatorname{min}}  \quad \| \bar{u}_{0}(t) - u_l(t) \|_{R}^{2} \\
    {s.t.} & \quad \begin{bmatrix}
    \bar{u}_{[-T_{\text{ini}} , N_p + T_{\text{ini}} -1]}(t) \\
    \bar{y}_{[-T_{\text{ini}} , N_p + T_{\text{ini}} -1]}(t)
    \end{bmatrix}
    =
    \begin{bmatrix}
    H_{L}(u^d) \\
    H_{L}({y}^d)
    \end{bmatrix}
    \alpha(t), \\
    & \quad \begin{bmatrix}
    \bar{u}_{[-T_{\text{ini}},-1]}(t) \\
    \bar{y}_{[-T_{\text{ini}},-1]}(t)
    \end{bmatrix}
    =
    \begin{bmatrix}
    u_{[t-T_{\text{ini}},t-1]} \\
    y_{[t-T_{\text{ini}},t-1]}
    \end{bmatrix}, \\
    & \quad \bar{u}_k(t) \in \mathcal{U}, \quad \forall k \in\{0, \ldots, N_{p}-1\}, \\
    & \quad \bar{y}_k(t) \in \mathcal{Y}, \quad \forall k \in\{0, \ldots, N_{p}-1\}, \\
      \bar{u}&_{k}(t)  , \bar{y}_{k}(t) \in \mathcal{S}_f, \quad \forall k \in \{N_p, \ldots, N_{p}+T_{\text{ini}}-1\}.
\end{align}
\end{subequations}

The optimization problem above must be solved at each time step $t$. In the Data-Driven Safety Filter detailed in algorithm \ref{algorithm}, this finite horizon optimization problem is implemented in a receding horizon manner. If this problem is feasible at $t=t_0$, then ${Traj}_{_{\text{ini}}}$ is in the safe set $S$ at $t_0$, and the solution $\bar{u}(t_0)$ provides the required safe backup trajectory. The recursive feasibility of the proposed DDSF is established in theorem (\ref{Safe:Theorem}), which in turn implies that the solution to this receding horizon algorithm ensures safety in the sense of definition (\ref{Safe: set}), i.e. it will keep the system safe as $t \to\infty$. It should be noted that both $\bar{u}$ and $\bar{y}$ are dependent on $\alpha$, and not independent variables \cite{9654975}. 

\begin{algorithm} [ht] \label{algorithm}
  \caption{Data-Driven Safety Filter}
  \label{DDSF: Algrithm}
  \begin{algorithmic}[1]
    \State Initialize $H_L(u^d)$, $H_L(y^d)$, $N_p$, $T_{\text{ini}}$, $Traj_{\text{ini}}$.
    \While{true} 
      \State Solve problem (\ref{equation: safety filter optimization problem}) for any (possibly unsafe) learning input.
      \State Apply $\bar{u}_0(t)$ to system (\ref{Discrete State Space}).
      \State Measure system's output, and update $Traj_{\text{ini}}$.      
      \If {the learning process is done}
        \State \textbf{break}
      \EndIf
      
    \EndWhile
  \end{algorithmic}
\end{algorithm}

\begin{Theorem}[Recursive Feasibility] \label{Safe:Theorem}
Let assumptions (\ref{Assumption: I}-\ref{Assumption: III}) hold, and $T_{\text{ini}} \geq l(A, C)$. The DDSF optimization problem (\ref{equation: safety filter optimization problem}) is feasible for all $t>t_0$, if it is feasible at $t = t_0$.
\end{Theorem}

\begin{proof}
Feasibility of optimization problem (\ref{equation: safety filter optimization problem}) results in the existence of a finite-length backup trajectory ($Traj_{p}$ and $Traj_{s}$) for the given initial trajectory $Traj_{ini}$ at $t=t_0$. The safe input $\bar{u}_{0}(t_0)$ is applied to the system (\ref{Discrete State Space}) at time $t = t_0$, and it will evolve to a new initial trajectory at $t=t_0+1$. This updated initial condition has at least one feasible solution since we calculated a feasible backup trajectory for it in the previous step. We can conclude that if the optimization problem (\ref{equation: safety filter optimization problem}) is feasible at $t_0$, it will also be feasible at $t_0+1$. Finally, by relying on induction, it follows that for any $t > t_0$, the problem (\ref{equation: safety filter optimization problem}) remains feasible.
\end{proof}

\begin{table}[ht]
  \centering
  \caption{System and Safety Filter Parameters  of Example 1}
  \label{table:I}
  \renewcommand{\arraystretch}{1.1} 
  \begin{tabular}{ |c|c|c|c| }
    \multicolumn{4}{l}{\textbf{Data-Driven Safety Filter}} \\
    \hline \hline
    \textbf{Parameter} & \textbf{Value} & \textbf{Parameter} & \textbf{Value} \\  
    \hline
    \(N_p\) & 20 & \(S_f\) & e.q. point of system (\ref{quad state space}) \\ 
    \hline
    \(R\) & $\mathbf{I}_{4 \times 4}$ & \(L\) & 24 \\ 
    \hline        
    \hline
  \end{tabular}
  \quad
  \begin{tabular}{ |c|c|c|c| }
    \multicolumn{4}{l}{\textbf{System's Properties}} \\
    \hline \hline
    \textbf{Parameter} & \textbf{Value} & \textbf{Parameter} & \textbf{Value} \\  
    \hline
    \(I_{xx}\) & \( 10^{-3} \, [\text{Kg.m}^2]\) & \(I_{yy}\) & \( 10^{-3} \, [\text{Kg.m}^2]\) \\ 
    \hline
    \(I_{zz}\) & \( 10^{-3} \, [\text{Kg.m}^2]\) & \textit{mass} & 0.2 [Kg] \\ 
    \hline
    \(g\) & \(9.81 \, \text{[m/$\text{sec}^2$]}\) & \(T_s\) & 0.1 [sec] \\
    \hline
    \(u_{1_{\text{min}}}\) & \(-1 \, \text{[N]}\) & \(u_{1_{\text{max}}}\) & \(+1 \, \text{[N]}\) \\ 
    \hline 
    \(u_{2_{\text{min}}}\) & \(-0.1 \, \text{[N.m]}\) & \(u_{2_{\text{max}}}\) & \(+0.1 \, \text{[N.m]}\) \\ 
    \hline 
    \(u_{3_{\text{min}}}\) & \(-0.1 \, \text{[N.m]}\) & \(u_{3_{\text{max}}}\) & \(+0.1 \, \text{[N.m]}\) \\ 
    \hline 
    \(u_{4_{\text{min}}}\) & \(-0.1 \, \text{[N.m]}\) & \(u_{4_{\text{max}}}\) & \(+0.1 \, \text{[N.m]}\) \\ 
    \hline
    \(\theta_{\text{min}}\) & \(-0.2 \, \text{[rad]}\) & \(\theta_{\text{max}}\) & \(+0.2 \, \text{[rad]}\) \\ 
    \hline
    \(\phi_{\text{min}}\) & \(-0.2 \, \text{[rad]}\) & \(\phi_{\text{max}}\) & \(+0.2\, \text{[rad]}\) \\ 
    \hline
    \(\psi_{\text{min}}\) & \(-0.2 \, \text{[rad]}\) & \(\psi_{\text{max}}\) & \(+0.2 \, \text{[rad]}\) \\ 
    \hline
    \(x_{\text{min}}\) & \(-1 \, \text{[m]}\) & \(x_{\text{max}}\) & \(+1 \, \text{[m]}\) \\ 
    \hline
    \(y_{\text{min}}\) & \(-1 \, \text{[m]}\) & \(y_{\text{max}}\) & \(+1 \, \text{[m]}\) \\ 
    \hline
    \(z_{\text{min}}\) & \(-1 \, \text{[m]}\) & \(z_{\text{max}}\) & \(+1 \, \text{[m]}\) \\ 
    \hline 
    \hline
  \end{tabular}
\end{table}

\section{Illustrative Examples}

To demonstrate the functionality of DDSF, we consider two linear systems in simulation, a six-degree-of-freedom (6-DOF) quadrotor representing a MIMO high-order system and a cruise control system representing a time-delay system.

\subsection{Example of a high-order system: 6-DOF quadrotor}

We use the linearized form of the 6-DOF quadrotor model due to its inherent instability and unsafe properties \cite{Sabatino860649}. The minimal state-space representation of $12$ states is given below.
\begin{equation} \label{quad state space}
    \dot{x} = A x + B u,
\end{equation}
where ${x}(t) \in \mathbb{R}^{12}$,  ${y}(t) \in \mathbb{R}^{6}$, and  ${u}(t) \in \mathbb{R}^{4}$ are state, output, and input vectors are represented as follows,
$$
\begin{array}{ccc}
    {x}=\left[\begin{array}{llllllllllll}
    \phi & \theta & \psi & p & q & r & u & v & w & x & y & z
    \end{array}\right]^T , \\[0.5ex]
    {u}=\left[\begin{array}{llll}
    u_1 & u_2 & u_3 & u_4
    \end{array}\right]^T , \\[0.5ex]
    {y}=\left[\begin{array}{llllll}
    \phi & \theta & \psi & x & y & z
    \end{array}\right]^T .
\end{array}
$$
Furthermore, $A$ and $B$ are sparse matrices, and their non-zeros elements are given as follows, where $A[i,j]$ is the $i^{th}$ row and $j^{th}$ column of matrix A,
\begin{align*}
    \begin{matrix}
        A[1,4] = 1 ,& A[2,5] = 1 ,& A[3,6] = 1 ,\\[0.5ex]
        A[10,7] = 1 ,& A[11,8] = 1 ,& A[12,9] = 1 ,\\[0.5ex]
        A[8,1] = g ,& A[7,2] = -g ,& B[9,1] = \frac{1}{mass} ,\\[0.5ex]
        B[4,2] =\frac{1}{I_{xx}} ,& B[5,3] =\frac{1}{I_{yy}} ,& B[6,4] =\frac{1}{I_{zz}}.
    \end{matrix}
\end{align*}
Moreover, we should note that all angular and transitional velocities are internal states that are not measured. All of the safety filter parameters and system properties are listed in Table  \ref{table:I}.

\begin{figure}[ht] 
  \centering
  \includegraphics[width=0.49\textwidth]{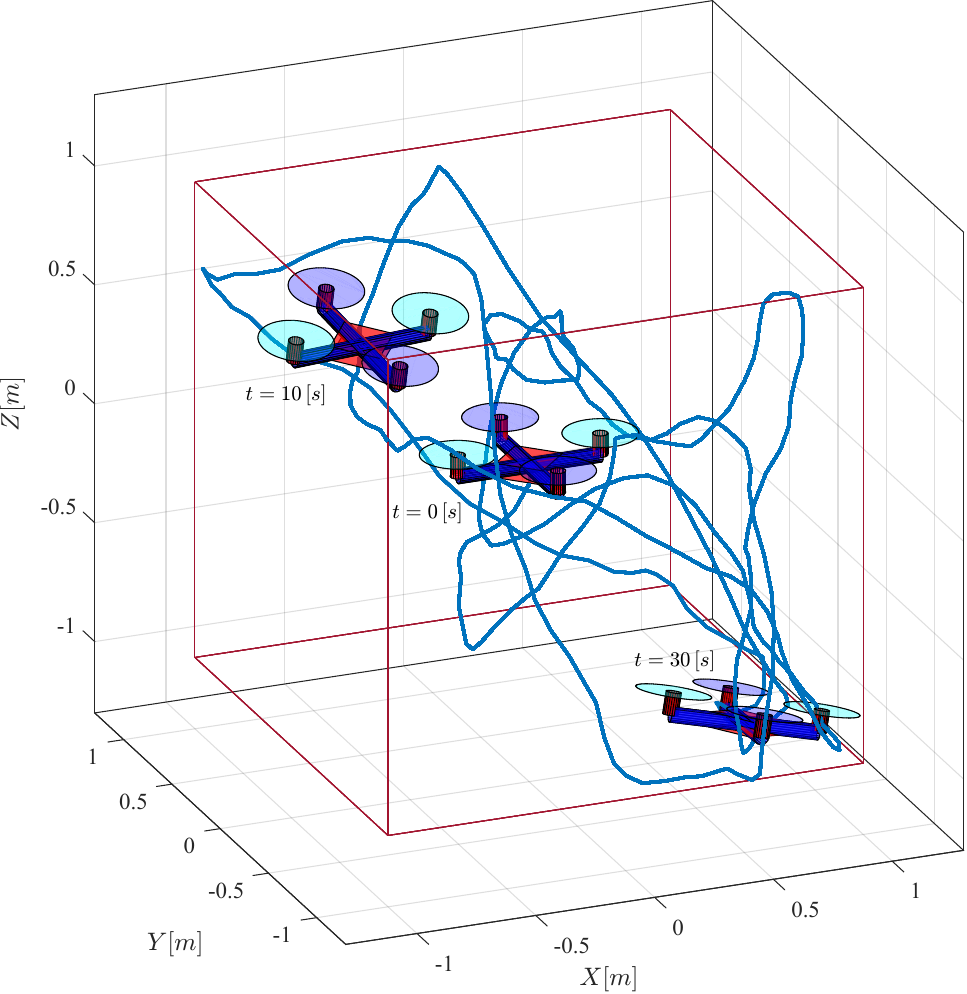}\vspace{-0.3cm}
  \caption{The evolution of the system's output trajectory in three-dimensional space under the influence of learning inputs for 30 seconds. DDSF keeps the system within predefined bounds, despite unsafe learning inputs (Example 1: 6-DOF quadrotor)}
  \label{fig: 3D_resuts}
\end{figure}

\begin{figure}[ht] 
  \centering
\includegraphics[width=0.49\textwidth]{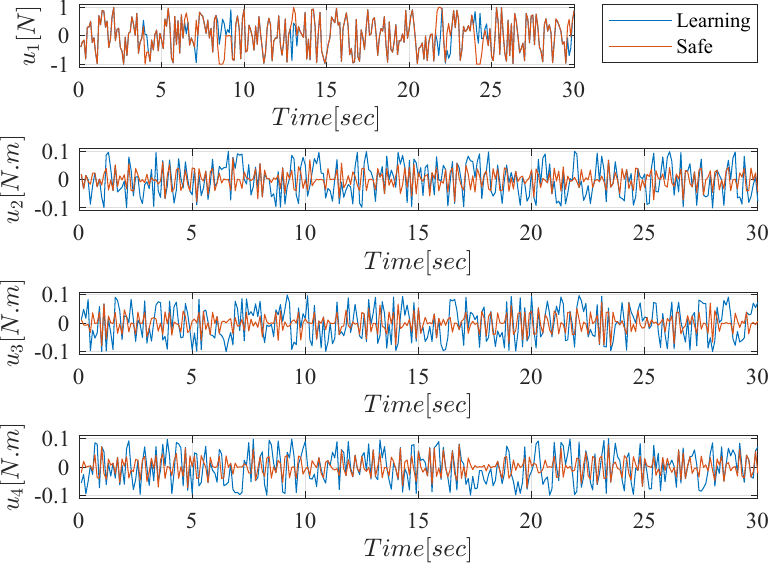}\vspace{-0.2cm}
  \caption{Learning inputs and corresponding modified safe inputs resulted by DDSF. (Example 1: 6-DOF quadrotor)}
  \label{inputs}
\end{figure}

\begin{figure}[ht] 
  \centering
  \includegraphics[width=0.49\textwidth]{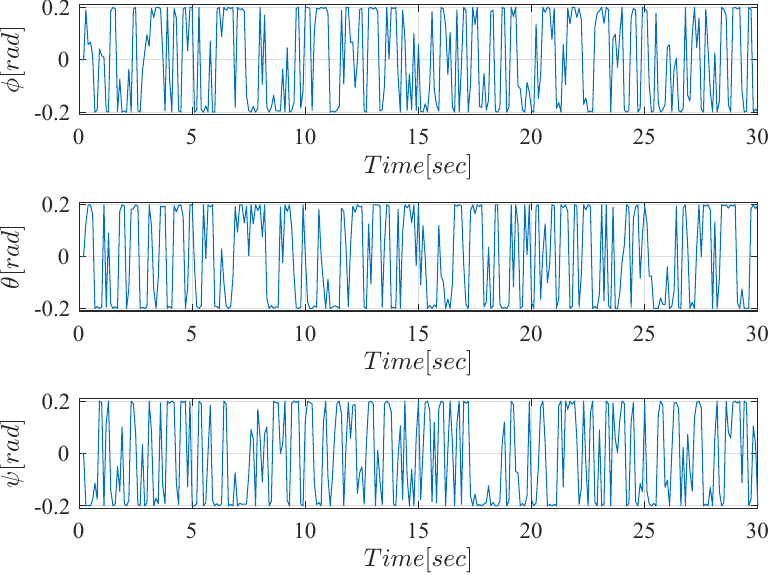}\vspace{-0.3cm}
  \caption{System's angle trajectories in the presence of learning inputs limited in $[-0.2,+0.2] \, [rad]$. (Example 1: 6-DOF quadrotor)} 
  \label{angles}
\end{figure}

\begin{figure}[t] 
  \centering
  \includegraphics[width=0.49\textwidth]{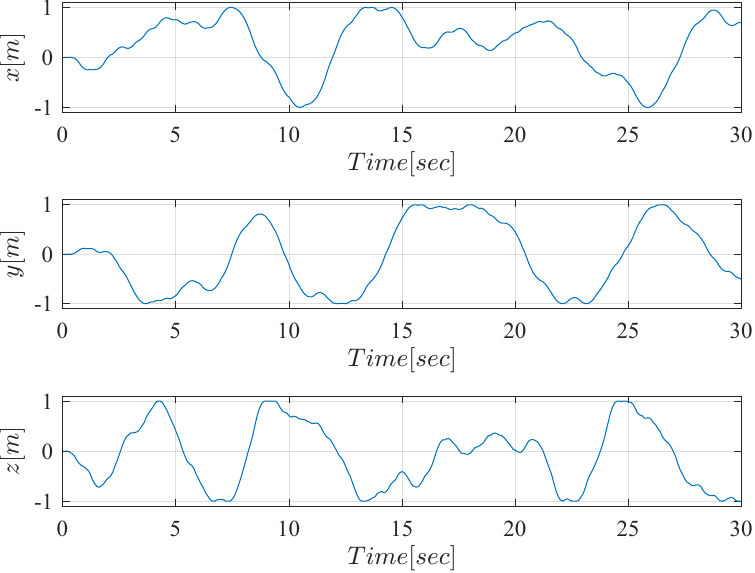}\vspace{-0.3cm}
  \caption{System's position trajectories in the presence of learning inputs limited in $[-1,+1] \, [m]$. (Example 1: 6-DOF quadrotor)}
  \label{position}
\end{figure}

Based on the definition of the system's lag (\ref{definition: lag}), $l(A, C) = 2$, any choice greater than this is acceptable for $T_{\text{ini}}$. Furthermore, we assume that our system's initial trajectory is at the equilibrium point where all input and output sequences are equal to zero to ensure the safe control problem is feasible at time $t = 0$. The learning input is a Pseudo-Random-Binary-Signal (PRBS) multiplied by a uniform random magnitude. The result of this scenario for $30 \,[sec]$ with the sampling time $T_s =0.1 \, [sec]$ is reported in Fig. \ref{fig: 3D_resuts}. The proposed DDSF ensures that the quadrotor position is bounded in the unit square despite the presented random inputs. As an example of DDSF functionality, consider the $z$-trajectory in Fig. \ref{position} and $u_1$-input in Fig. \ref{angles} between $10\text{ and }15 \, [sec]$. There is a considerable deviation between learning and safe input as the $z$ trajectory approaches the boundary of the defined safety bounds. To prevent such an unsafe scenario, the DDSF increased the thrust of all motors, $u_1$, to prevent the system from collapsing. Furthermore, Euler angles, positions, and control inputs are all limited in their predefined range and illustrated by Fig. \ref{inputs}-\ref{position}, respectively.

\vspace{-0.2cm}
\begin{figure}[H] 
  \centering
  \includegraphics[width=0.49\textwidth]{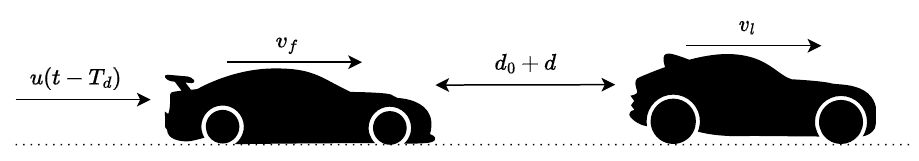}\vspace{-0.1cm}
  \caption{ A graphical illustration of the time-delay adaptive cruise control system (Example 2: System with unknown time delay)}
  \label{Two_car_figure}
\end{figure}

\subsection{Example of system with unknown time delay: adaptive cruise control system}

\begin{figure*}[ht] 
  \centering
  \includegraphics[width=0.99\textwidth]{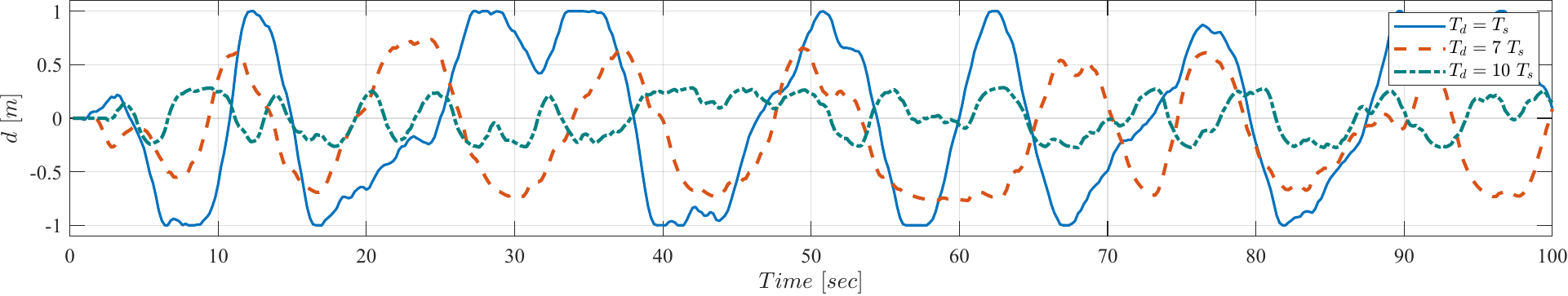}
  \caption{ Distance variation between two cars considering three different unknown input time delays. (Example 2: System with unknown time delay)}
  \label{distance}
\end{figure*}

As a second example, we use a modified model of Adaptive Cruise Control (ACC) given in \cite{taylor2020adaptive}. It represents the distance between a follower and a lead car. Let the leader cars' velocity, $v_l$, be constant, then relative distance considering input-delay, $T_d$, is defined as follows,
\begin{equation}
\left[\begin{array}{l}
    \dot{d} \\
    \ddot{d}
    \end{array}\right] = \left[\begin{array}{cc}
     0& 1 \\
     0& 0
    \end{array}\right] \left[\begin{array}{l}
    d \\
    \dot{d}
    \end{array}\right] + \left[\begin{array}{c}
    0 \\
    \frac{1}{m_{c}}
    \end{array}\right]  u (t-T_d),
\end{equation}
where $d \ [m]$ is the distance variation between two cars limited in $[-1,+1]$, $\dot{d} = v_l -v_f\ [\frac{m}{s}]$ is the relative velocity, $m_{c} = 1650 \ [Kg]$ is the follower car's mass, and $u \ [N]$ is the follower car's control input limited in $[-2000,+2000]$. Assuming sampling time $T_s = 0.2 \ [sec]$, prediction horizon $N_p = 15$, and duration of initial trajectory $T_{\text{ini}} = 15$ steps, we evaluate the effectiveness of DDSF for (unknown) time delays $T_d \in \{ T_s, 2 T_s, 3 T_s, \hdots, 10 T_s\}$. A graphical illustration of this system is given in Fig. \ref{Two_car_figure}. We consider $d$ to be the only measurable output. Similar to the previous example, a random signal is chosen as the learning input. The result of the simulation in terms of distance variation with respect to time is shown in Fig. \ref{distance}. As it is clear, when the dead time is small, $T_d = T_s$, we can still cover the whole of the output admissible set. When the time delay is increased, $T_d = 7 T_s$ or $T_d = 10 T_s$, the input-output safe set is reduced as expected, and the safety filter acts more conservatively. Note that the DDSF does not use explicit information on the time delay, and this result follows entirely from the information in the data.

\section {Discussion and Concluding Remarks}

This paper introduced the Data-Driven Safety Filter (DDSF) that utilizes pre-recorded system trajectories to prevent unsafe behaviors. Two simulation examples, a high-order system, and a time-delay system, have been presented to highlight the effectiveness of the method. One of the key features of the presented method is the ability to maintain the system inside a safe set solely by relying on input-output measurements. DDSF can offer a safe learning process in conjunction with any data-driven, learning-based, or Human-in-the-loop algorithm, directly from data. The proposed DDSF is presented in a deterministic framework. In real-world applications, disturbances, noise, and nonlinear behavior can lead to prediction errors and constraint violations. Robust solutions to these challenges are typically obtained by tightening the constraints, which introduces conservatism. 

As there is a similarity between the defined input-output safe set in definition (\ref{Safe: set}) and the region of convergence in MPC, conservatism introduced by the finite horizon approximation and the choice of the terminal invariant set can be reduced by increasing the prediction horizon or by substituting the equilibrium point invariant set definition with a more extensive invariant set. DDSF is a quadratic programming problem that can be solved efficiently. However, when long prediction horizons are considered, the DDSF formulation becomes computationally expensive to solve. The equivalence of specific data-driven solutions with explicit models can be exploited to formulate an equivalent input-output safety filter that is less computationally expensive.

\bibliographystyle{IEEEtran}
\balance
\bibliography{root.bib}

\end{document}